\RequirePackage{fix-cm}
\documentclass[twocolumn]{svjour3}          
\smartqed  
\usepackage{graphicx}
\usepackage{lipsum}
\usepackage{amsmath}
\usepackage{microtype}
\usepackage{enumerate}
\usepackage{subcaption}
\usepackage{setspace}
\usepackage{xcolor}
\definecolor{blue1}{rgb}{0,0.5,0.7}

\journalname{Nature Applied Sciences}

\begin{document}

\title{The recent advances in the mathematical modelling of human pluripotent stem cells 
}

\titlerunning{Mathematical modelling of hESCs}        

\author{L E Wadkin$^{*1}$ \and S Orozco-Fuentes$^1$ \and I Neganova$^2$ \and  M Lako$^3$ \\\and A Shukurov$^1$ \and N G Parker$^1$}


\institute{              *\email{l.e.wadkin@ncl.ac.uk}           
[1] School of Mathematics Statistics and Physics, Newcastle University, UK, [2] Institute of Cytology, RAS St Petersburg, Russia, [3] Institute of Genetic Medicine, Newcastle Univeristy, UK. 
}
\date{Received: date / Accepted: date}

\maketitle

\begin{abstract}
Human pluripotent stem cells hold great promise for developments in regenerative medicine and drug design. The mathematical modelling of stem cells and their properties is necessary to understand and quantify key behaviours and develop non-invasive prognostic modelling tools to assist in the optimisation of laboratory experiments. Here, the recent advances in the mathematical modelling of hPSCs are discussed, including cell kinematics, cell proliferation and colony formation, and pluripotency and differentiation. 
\end{abstract}

\section{Introduction}
\label{intro}
\begin{figure}
	\includegraphics[width=1\columnwidth]{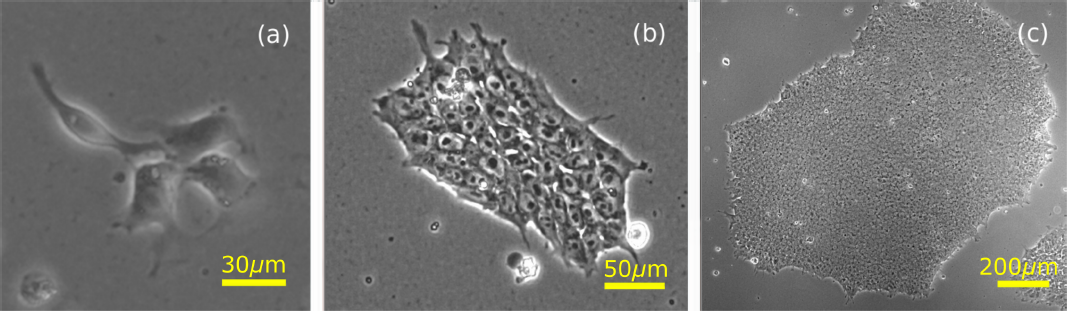}
	\caption{Microscopy images of hESCs showing growing colonies from (a) a few cells up to colonies of (b) hundreds and (c) thousands.}
	\label{fig:colonies}       
\end{figure}

Human pluripotent stem cells (hPSCs) have the ability to self-renew indefinitely through repeated divisions (\textit{mitosis}) and can \textit{differentiate} into any bodily cell type (the \textit{pluripotency} property). The latter property underpins their promising clinical applications in drug discovery, cell-based therapies and personalised medicine \cite{Ebert,Zhu2013}. Amongst others, cardiomyocytes  \cite{heartcells}, pancreatic cells \cite{panc} and corneal cells \cite{cornea} have all been successfully created from hPSCs. In the lab, hPSCs are grown in mono-layer colonies of up to thousands of cells (Figure~\ref{fig:colonies}) from which they can be directed for specific experiments or therapies, or expanded to produce further hPSC colonies. They occur either as human embryonic stem cells (hESCs) derived from the early embryo, or human induced pluripotent stem cells (hiPSCs) which are derived by the genetic reprogramming of differentiated cells \cite{ipscs}. The latter approach, which received the 2012 Nobel Prize in Medicine or Physiology for its discovery, offer patient-specific hPSCs without the ethical issues associated with hESCs.

Emerging biomedical technologies require the efficient, large-scale production of hPSCs \cite{KROPP2017244}. Furthermore, applications of hPSCs in the clinic require great control over the pluripotency, \textit{clonality} (the proportion of identical cells that share a common ancestry) and differentiation trajectories \textit{in-vitro}.  However, the existing procedures for large scale experiments remain inefficient and expensive due to low cloning efficiencies of 1\% to 27\% (the percentage of single cells seeded that form a clone) \cite{Andrews,ROCK1}. Understanding factors which promote the efficient generation and satisfactory control of hPSC colonies (and their derivatives) is a key challenge. 

Mathematical and computational modelling allows the identification of generic behaviours, providing a framework for rigorous characterisation, prediction of observations, and a deeper understanding of the under-lying natural processes. The application of mathematics to biology \cite{Murray2002} has led to many significant achievements in medicine and epidemiology (for example, predicting the spread of `mad cow' disease \cite{BARNES201385,bse} and influenza \cite{KISSLER201986}), evolutionary biology \cite{evo} and cellular biology (descriptions of chemotaxis \cite{chemotaxis} and predicting cancer tumour growth \cite{Altrock}). Similarly, mathematical models are a powerful tool to further our understanding of hPSC behaviours and optimise crucial experiments.

The first mathematical model of stem cells, a stochastic model of cell fate decisions \cite{Till64}, has since been extended to include many other aspects of cell behaviour \cite{ppmodels,Olariu,Xu,Hannam_2017,MacLean17}. In particular, when such mathematical models are rigorously underpinned and validated on experimental observations, the reciprocal benefit for experimentation can be profound: an example is the development of an experimentally-rained model of hiPSC programming, which led in turn to strategies for marked improvements in reprogramming efficiency \cite{Hanna09}.
	
Coherent mathematical models of hPSC properties may provide non-invasive prognostic modelling tools to assist in the optimisation of laboratory experiments for the efficient generation of hPSC colonies. Statistical analysis of experimental data allows the quantification of stem cell behaviour which can then inform the development of these models. Here we shall discuss recent advances in the mathematical modelling of hPSCs and their impact.

This review focuses mostly on hESCs, with some limited discussion of hiPSCs. We first outline some of the key properties of hPSCs before focussing on recent developments in mathematical models of the key properties:
\begin{itemize}
	\item \textit{Section~\ref{sec:properties}: Key biological properties of hPSCs}

	\item \textit{Section~\ref{sec:kin}: Cell kinematics}. The movement of cells alone, in relation to one another and within hPSC colonies.
	\item \textit{Section~\ref{sec:prolif}: Colony growth}. Models capturing cell proliferation, with and without a spatial component. 
	\item \textit{Section~\ref{sec:cellpp}: Cell pluripotency}. Pluripotency regulation models, both intra-cellular and at the colony scale. 
\end{itemize}
Finally, in Section~\ref{sec:disc} we provide a summary of the models discussed, their impact on biological experiments and the next steps for model development.

\section{Key biological properties of hPSCs}
\label{sec:properties}

\begin{figure}
	\includegraphics[width=1\columnwidth]{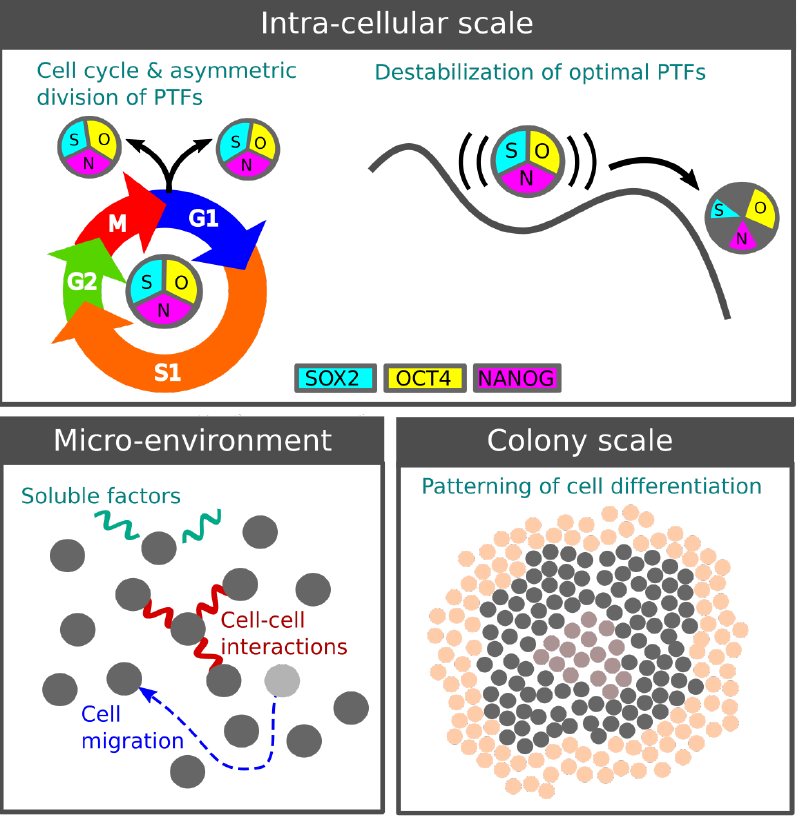}
	\caption{Scales of hPSC behaviour: (a) Intra-cellular scale	e.g., cell cycle,  division, inheritance of pluripotency factors, PTF. (b) Cell micro-environment e.g., interaction with other cells, the medium and substrate. (c) Colony-scale phenomena e.g., patterning of differentiated cells.}
	\label{fig:cellscales}       
\end{figure}

The satisfactory understanding and control of hPSC evolution remains elusive due to their complex behaviour over multiple scales: the intra-cellular scale (processes happening within cells), the cellular or micro-environment scale (the environmental effects on individual cells) and the colony scale (collective cell behaviours throughout colonies), as illustrated in Figure~\ref{fig:cellscales}. Advances in imaging and molecular profiling (classification based on gene expression) have identified the core processes within the evolving colony \cite{Andrews,Tokunaga,Maddah,Suga}. Here we outline some of these key biological properties across these scales and their relevance for mathematical modelling.

\subsection{Intra-cellular scale}

The key intra-cellular behaviours integral to hPSC modelling are the cell cycle and pluripotency regulation. The cell cycle is the timed series of events controlling DNA replication and resulting in a cell division. The phases of the cell cycle are: G1 (growth phase), S (synthesis phase in which DNA is replicated), G2 (further growth) and M (mitosis, the cell division). The G1 phase is shortened for hPSCs, leading to more rapid proliferation than for somatic cells \cite{shortg1}.

The maintenance of pluripotency depends on the stable inter-regulation of pluripotency transcription factors (PTFs) \cite{ppgenenetwork}, mainly by the genes OCT4, SOX2 and NANOG \cite{Zhang67}. Fluctuations of the PTF abundances are believed to cause the variation in pluripotency in different sub-populations \cite{ppgenenetwork}. Destabilisation and the interaction of these PTFs with chemical signalling pathways triggers \textit{differentiation}, the departure from the pluripotent state \cite{ppgenenetwork,Kumar14} towards specific cell fates \cite{WANG2012440}. The cell cycle also affects pluripotency  and cell fate \cite{PAUKLIN2013135} and vice versa \cite{Zhang67,Ouyang,Lee171}. Moreover, recent work suggests that the PTFs are inherited asymmetrically as a cell divides \cite{Wolff}, biasing the fate of the daughter cells and contributing to colony heterogeneity. 

\subsection{Micro-environment}
As in the embryo, the local environment of the cell is key to its \textit{in-vitro} evolution. One of the leading environmental factors affecting hPSCs is the substrate on which they are grown.  Substrates may either consist of a layer of mouse or human `feeder' cells or a protein substrate, with the latter growing in popularity for clinical application since they avoid the risk of genetic contamination. The substrate influences pluripotency \cite{HWANG2008} and mobility \cite{Zaman10889} through its growth factors and adhesion forces. Low cell motility improves clonality by suppressing cross-contamination of colonies \cite{Chang2019}, although its role in colony heterogeneity is yet to be established. 

As well as the substrate, cell-cell interactions are also important. hPSCs benefit from being in colonies where they exhibit higher viability and pluripotency \cite{Chen10}. hPSCs apparently sense each other up to a distance of around $150\,\mu$m (of order 5 cell diameters) \cite{Li,me2}. Meanwhile, as the colony grows and becomes denser, the mutual mechanical pressure of the hPSCs can affect the cell cycle \cite{Wu15}.

\subsection{Colony scale}
Perhaps most intriguing, yet least understood, are behaviours that emerge on a colony scale. The promotion of pluripotency in larger colonies \cite{Bauwens08,Nemashkalo3042} shows that single cells are influenced by the whole colony. Indeed, it has been suggested that pluripotency is a collective statistical property of cells \cite{MACARTHUR13}, rather than a well-defined property of individual cells.

Further colony-scale effects are evident in the spatial patterning of the cell fates after differentiation. Mechanical forces and chemical signals operating over distances larger than the cell separation influences single-cell genetic expression to form bands of differentiated cells \cite{band} (illustrated in Figure~\ref{fig:cellscales}); these structures are enhanced under imposed boundaries, emphasizing the role of mechanical forces \cite{Warmflash14,Etoc16}. With further understanding, mechanical effects and boundaries could be harnessed to engineer
specific desired differentiated cells \cite{Xue18}.\\

Incorporating these complex behaviours over multiple scales into mathematical models is challenging. A key goal is to develop coherent models which capture the individual cell behaviours, e.g., cell kinematics and the inter-cellular maintenance of pluripotency, and lead to the observed collective effects on the colony scale, e.g., collective migration and the spatial patterning of pluripotency and differentiation.

\section{Cell kinematics}
\label{sec:kin}

Motility is an intrinsic property of hPSCs; they can increase their migratory activity under certain conditions \cite{migration1}. Their migration is achieved through adaptations in cell morphology via the reorganisation of the actin cytoskeleton to form a leading edge pseudopodia \cite{adherentcells}. Unregulated cell migration \textit{in-vitro} can cause clonality loss as the cell population grows, undesirable when a genetically identical clonal population is required \cite{clonality1,clonality2}. Furthermore, anomalous cell migration has been linked to deviations in the undifferentiated state of hiPSCs \cite{SHUZUI2019246}. A thorough understanding of the migration of hESCs is needed to optimise \textit{in-vitro} clonality and facilitate the development of therapies for migration related disorders. Here we discuss the kinematics of isolated cells and their pairs as well as cell migration within colonies.

\subsection{Kinematics of isolated cells and pairs}

hPSCs are often seeded at low density to preserve the clonal purity of the emerging colonies. Migration of individual cells between the incipient colonies can result in clonality loss. It is important therefore to quantify the migration of individual cells upon a growth plate.

The unconstrained motion of cells on a 2D plane can often be described as a 2D random walk, the simplest being Brownian motion \cite{codling,deBack19}. Random walks can be biased by an external source giving preference to movement in a particular direction (a biased random walk or BRW). A correlated random walk (CRW) involves a correlation in the direction of the next step in relation to the previous step, i.e., persistence, where the next step is more likely to be in the direction of the previous step, or anti-persistence, where the next step is more likely to be in the opposite direction. CRWs often occur in cell kinematics in the absence of external biases \cite{CRWamoeboid,CRWepithelial,CRWfibroblasts}. 

The diffusive nature of a random walk can be quantified by considering the mean square displacement (MSD) of cell trajectories. The MSD is a measure of the trajectory of a particle from its starting position over time, $\langle r^2 \rangle= \langle(\vec{x}-\vec{x_0})^2\rangle$ where $\vec{x}(t)$ is the position of the particle, $\vec{x}_0$ is the initial position at $t=0$ and angular brackets denote the average taken over all trajectories. For a typical diffusive particle, the MSD increases linearly with time, $\langle r^2 \rangle \propto D t$, where $D$ is the diffusion coefficient. The root mean square displacement is given by ${\langle r^2 \rangle}^{1/2}=\sqrt{2Dt}$, from which $D$ can be calculated. If $\langle r^2 \rangle \propto D t^\alpha$, with $\alpha<1$ the motion is sub-diffusive or super-diffusive with $\alpha>1$. 

\begin{figure}[h]
	\centering
	\begin{subfigure}[t]{0.49\columnwidth}
		\caption{}
		\includegraphics[width=\columnwidth]{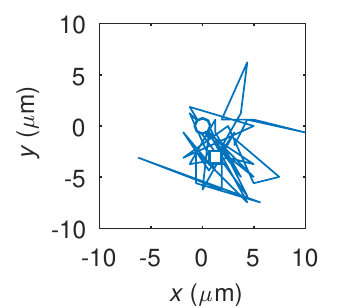}
		\label{fig:traj17}
	\end{subfigure}
	\begin{subfigure}[t]{0.49\columnwidth}
		\caption{}
		\includegraphics[width=\columnwidth]{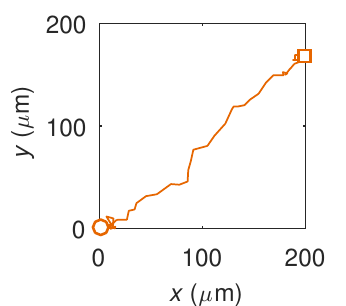}
		\label{fig:traj3}
	\end{subfigure}
	\\
	\begin{subfigure}[t]{1\columnwidth}
		\caption{}
		\includegraphics[width=\columnwidth]{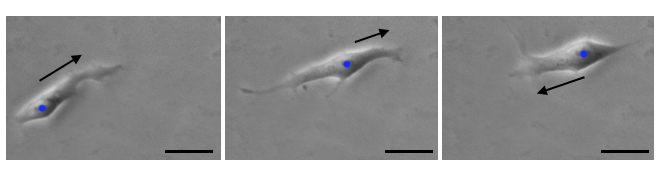}
		\label{fig:bf}
	\end{subfigure}
	\caption{\label{fig:traj}Example single cell trajectories for (\subref{fig:traj17}) isotropic motion around a central point and (\subref{fig:traj3}) a directed walk. The initial and final cell centroid positions are shown as a circle and a square respectively (note that these points are not representative of cell or nucleus size). (\subref{fig:bf}) A single hESC migrating backwards and forwards along a local axis. The blue dot shows the cell nucleus and the black arrow the direction of instantaneous velocity. The scale bars are $30\,\mu$m in length \cite{me2}.}
\end{figure}

The nature of individual cell movements has been observed through direct experiments with hPSCs (in particular hESCs) and analysed within the random walk framework \cite{Li,me2,me}. The movement of single hESCs has been described as an isotropic random walk when the cells are in isolation, i.e., more than approximately $150\,\mu$m away from any neighbouring cells. As the separation distance decreases the cell movements become more directed towards each other, with motility-induced re-aggregation occurring in 70\% of instances when the distance been two hESCs is less than $6.4\,\mu$m \cite{Li}. A minority of isolated single cells exhibit super-diffusive behaviour, contributing heavily to the motility related clonality loss \cite{Andrews,Li,me}. Example experimental trajectories for cells exhibiting typical diffusive behaviour and super-diffusive migration are shown in Figure~\ref{fig:traj}.

Our study containing further experimental analysis of hESCs \cite{me2} has shown evidence of correlated random walks of individual isolated stem cells. Single hESCs (more than $150\,\mu$m away from any neighbouring cells, as in \cite{Li}) tend to perform a locally anisotopic walk, moving backwards and forwards along a preferred local direction correlated over a time scale of around 50\,minutes, becoming more persistent over time. The motion is also aligned with the axis of cell elongation (Figure~\ref{fig:traj}) which could suggest an attempt to locate other neighbouring cells. Further experiments should quantify how the presence of multiple neighbours affects this anisotropic movement.

Our study also found that pairs of hESCs in close proximity tend to move in the same direction, with the average separation of $70\,\mu$m or less and a correlation length (the length scale of communication) of around 25\,$\mu$m. Often the pairs of cells remained connected by their pseudopodia, even at larger distances ($>100\,\mu$m) when they exhibited independent movements. For the correlated pairs, it is not known whether the movement correlation is facilitated by the physical connection or the coordination is due to cell-cell chemical contact alone.

There is evidence that cell migration in 3D does not follow a persistent random walk and new models will need  to be developed to accurately describe this motion \cite{Wu14}. These experimental results further inform the development of individual based models for cell migration as a random walk and can be integrated into more complex models of cell movement within colonies.

\subsection{Colony kinematics}

Stem cells also exhibit motion as part of larger groups and colonies. The coordinated migration of large numbers of hPSCs \textit{in-vivo} is essential in tissue generation \cite{MUGURUMA2015537} and wound healing \cite{pmid30602767}. The modelling of such larger groups and colonies of hPSCs is more complex, as both collective and individual behavioural effects are involved \cite{pmid24186932}. 

Popular agent-based models have been developed to incorporate these results into colony models, but the challenges still remain to fully capture the experimental behaviours, especially collective aspects and cell migration in 3D. These agent-based migration models are often combined with models of colony growth and proliferation \cite{NGUYEN2019625,Hoffman}.

hPSCs show coordinated intra-colony movements which cease upon differentiation \cite{ZANGLE2013593}. Cell movement speed varies within colonies, with higher average speed at the periphery and lower in the central region \cite{SHUZUI2019246}. Recently, a two-dimensional individual-based stochastic model was developed of cell migration, cell-cell connections and cell-substrate connections and captures well these experimental observations \cite{NGUYEN2019625}. The model introduces the energies of cell-cell and cell-substrate connections. Any energy released by breaking and forming these connections allows cell migration to one of the eight directions on a square lattice. The direction of movement is determined at random based on a probability related to the cell's energy and a spatial weighting which favours a side rather than a diagonal direction (as described in \cite{KINOOKA2000285}). Cell proliferation and quiescence (the reversible state of a cell in which it does not divide) are also included. The model suggests that cell division is a leading factor in the increased mobility at the colony edges, and will be useful for studying behaviours of hiPSCs and improving experiments.

\begin{figure}
	\includegraphics[width=1\columnwidth]{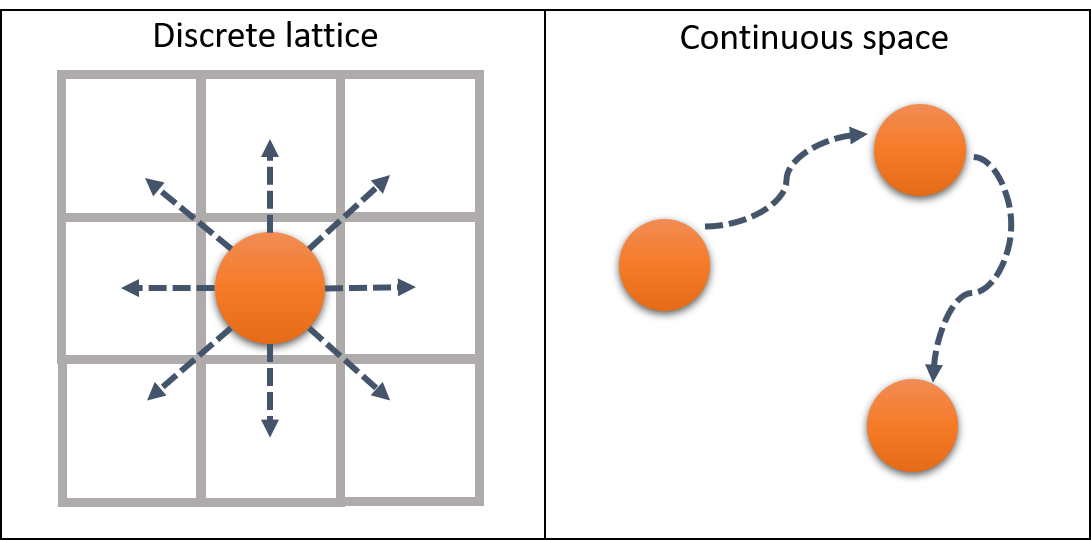}
	\caption{The migration of cells can be modelled either on a lattice or in continuous space.}
	\label{fig:lattice}       
\end{figure}

Modelling cell movement on a discrete lattice is widely used, e.g., for mesenchymal stem cell tissue differentiation \cite{Khayyeri} and cancer stem cell driven tumour growth \cite{abmcancer}.  Some models allow many lattice nodes per cell as in the Potts model \cite{cellularpotts}. There is also a range of agent-based continuous models where cell movement is not restricted to a grid but a cell can move continuously in any direction as illustrated in Figure~\ref{fig:lattice} \cite{offlattice,adra10}. Here the movement is described using forces or potentials with positions obtained from differential equations of motion for each cell. In centre based models (CBM), each cell is represented by a simple geometrical object, such as a circle, whereas in vertex models a cell is defined by a number of connected nodes \cite{VanLiedekerke2015}. These models will be discussed in more detail in Section~\ref{sec:prolif}.

There are also models which focus on the cells' changing morphology. For example, a model has been developed for mesenchymal stem cells which includes the random formation, elongation and retraction of pseudopodia, resulting in dragging forces which lead to cell movement \cite{Hoffman}. However, the model of Ref. \cite{Hoffman} shows more ballistic and accelerated dynamics than experimental results \cite{Dieterich459}.

\section{Colony growth}
\label{sec:prolif}

Colonies of hPSCs are formed by repeated mitosis in which two genetically identical daughter cells are produced from the division of the mother cell. The cell cycle is the sequence of events that occur in a cell in preparation for the division as described in Section 2.1. The simplest mathematical models incorporate cell proliferation probabilistically, with the division time for each cell drawn at random from a suitable probability distribution \cite{NGUYEN2019625}. Others go a step further by moving cells through each cell cycle phase according to timings based on experimental data \cite{WALKER200489} or as cell volume increases \cite{Hoffman}. Sometimes divisions do not occur; this probabilistic nature of self-renewal can be incorporated when the end of the cell cycle is reached \cite{Ganguly06}. There are also more complex models which describe the relationship between inter-cellular processes based on growth factors (proteins that regulate cell growth) \cite{adra10} and more sophisticated mathematical models describing the cell cycle in terms of limit cycles \cite{ADIMY20061091}. 
 
The doubling time of stem
cells number varies and can be affected by various environmental and chemical factors, including cell density and the colony maturity \cite{Andrews,Park08,Claassen09,Turner14}. Models of colony growth can be dynamical-system type models that address the time evolution of the colony size, or spatial models which track individual cells and the growing colony in space and time.

\subsection{Population dynamics models}

Population models have been used to understand the process by which blood cells are formed \cite{MacLean17}, cancer tumours grow \cite{Michor08} and the impact of hPSC colony growth on clonality \cite{meclonality}. Early population dynamics models for stem cells were based on stochastic birth-death processes \cite{Till64} involving systems of ordinary differential equations \cite{Loeffler80}. One of the most popular models for hPSCs includes two populations of dividing and non-dividing cells, with a term for accounting for cell loss through death or differentiation (often referred to as the Deasy model, which is a development of the Sherley model to include cell loss) \cite{Deasy03,Sherley}. The evolving number of cells over time $N(t)$ is obtained as
\begin{equation}
N(t)=N_0\left[\frac{1}{2}+ \frac{1-{(2\alpha)}^{t/D_t+1}}{2(1-2\alpha)}\right]-M,
\label{eq:Deasy}
\end{equation}
\noindent
where $N_0$ is the initial number of cells, $\alpha$ is the mitotic fraction, $D_t$ is the cell division time, and $M$ is the number of lost cells.

More recently, hyperbolastic growth models (a new class of parameter model for self-limited growth behaviours \cite{Tabatabai05}) have been introduced for both adult and embryonic stem cells \cite{Tabatabai2011}. These growth models provide more flexibility in the growth rate as the population reaches its carrying capacity and have been demonstrated to capture experimental data well \cite{Tabatabai05,Tabatabai2011}. The population in this case is governed by a non-linear differential equation
\begin{equation}
\frac{dN(t)}{dt}=(L-N(t))\left[ \delta \gamma t^{\gamma-1}+\frac{\theta}{\sqrt{1+\theta^2t^2}}\right],
\label{eq:H3}
\end{equation}
\noindent
with the initial condition $N(0)=N_0$, and the parameters $L$ (representing the limiting value, or carrying capacity of the population), $\delta$ (the intrinsic growth rate), $\gamma$ (a dimensionless allometric constant) and $\theta$ (additional term allowing for the variation in the growth rate). This model can be used to describe both proliferation and cell death rates more accurately than Equation~(\ref{eq:Deasy}) \cite{Tabatabai2011}. 

\begin{figure}[h]
	\centering
	\begin{subfigure}[t]{0.59\columnwidth}
		\caption{}
		\includegraphics[width=\columnwidth]{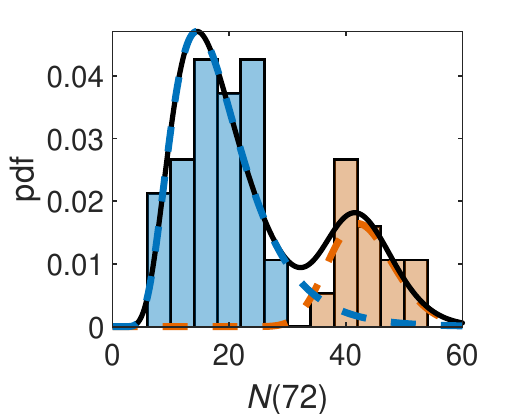}
		\label{fig:n72}
	\end{subfigure}
	\begin{subfigure}[t]{0.39\columnwidth}
		\caption{}
		\includegraphics[width=\columnwidth]{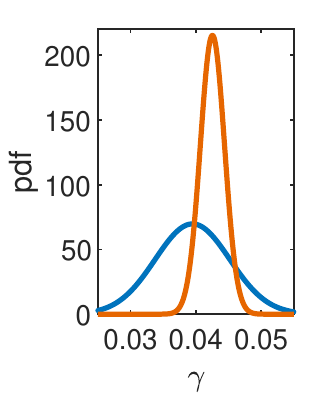}
		\label{fig:gamma}
	\end{subfigure}
	\caption{\label{fig:N72}(\subref{fig:n72}) The colony populations at 72\,h after seeding with a lognormal mixture model fitting for the single founding cell population (blue) and the pair founding cell population (orange). (\subref{fig:gamma}) The growth rate probability distributions for both populations. Adapted from \cite{meclonality}.} 
\end{figure}

Our most recent work develops a population model of the growth for hESC colonies based on experimental data \cite{meclonality}. We analysed the evolution of the colony populations and found that the distribution of colony sizes was multi-modal, corresponding to colonies formed from a single cell and colonies formed from pairs of cells as shown in Figure~\ref{fig:N72}. The colony populations can be described using a stochastic exponential growth model, with the growth rates of colonies emerging from single cell and cell pairs being drawn from normal distributions: 
\begin{equation}
\begin{cases}
N_\textrm{A}=e^{\gamma_{\rm A} t}, \,\, \gamma_{\rm A}\sim \rm{N}(\mu_{A}, {\sigma_{A}}^2),&\text{probability $\alpha$},\\

N_\textrm{B}=2e^{\gamma_{\rm B} t}, \,\, \gamma_{\rm B}\sim \rm{N}(\mu_{B}, {\sigma_{B}}^2),&\text{probability $\beta$,}
\end{cases}
\label{eq:model2}
\end{equation}
\noindent
with $\mu_{\rm A}=0.039$ and $\sigma^2_{\rm A}=0.006^2$, $\mu_{\rm B}=0.043$, $\sigma^2_{\rm B}=0.002^2$, $\alpha=0.77$ and $\beta=0.23$ inferred from the fitting to the experimental data shown in Figure~\ref{fig:N72}. The growth rate for colonies emerging from pairs of cells is greater than for colonies founded by single cells. This means that colonies that have grown from cell pairs are larger not only due to the initial condition but also because their proliferation rate is larger. This is consistent with observations that hPSCs proliferate more effectively when in close proximity to other cells \cite{Chen10,Moogk}. This difference is important when the clonality of a colony needs to be assessed non-invasively, e.g., from its size.

The model can be used to predict hPSC colony growth and to calculate the time scales over which colony size no longer predicts the number of founding cells based on their seeding density. This model can also be used to simulate colony growth in space which is discussed in the next section.

\subsection{Spatial modelling}

Colony growth can also be modelled spatially and, as with cell migration, the models can either be set on a regular or irregular lattice or in continuous space. Each cell can be modelled individually in an agent-based model, or for large numbers of cells where agent-based models become computationally challenging, using continuum models. A thorough summary of these different model types, along with their advantages and disadvantages with a view to tissue mechanics is provided in \cite{VanLiedekerke2015}. Here the recent attempts to model hPSC colonies using a variety of these techniques will be discussed. 

\begin{figure}[h]
	\centering
	\begin{subfigure}[t]{0.55\columnwidth}
		\caption{}
		\includegraphics[width=\columnwidth]{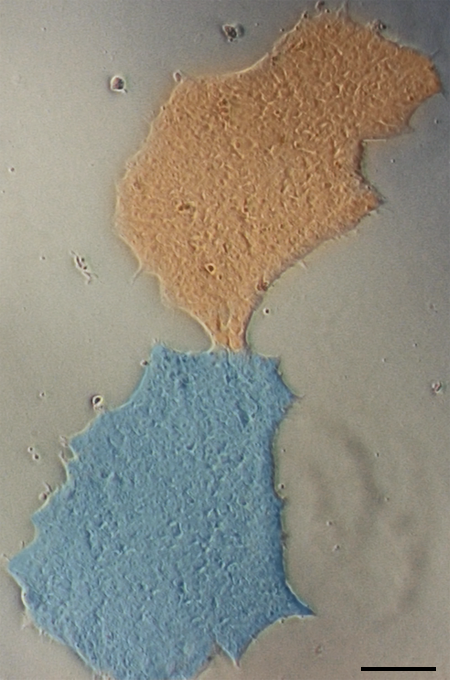}
		\label{fig:mergea}
	\end{subfigure}
	\begin{subfigure}[t]{0.43\columnwidth}
		\caption{}
		\includegraphics[width=\columnwidth]{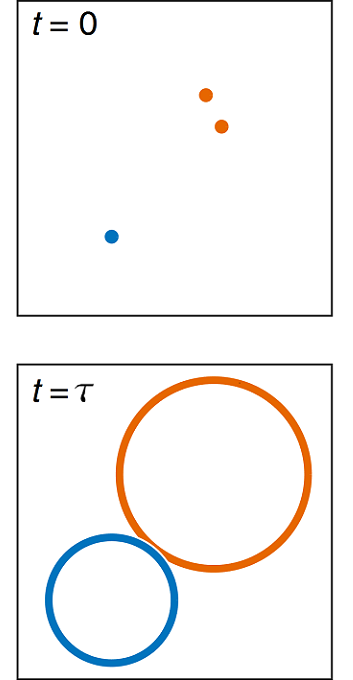}
		\label{fig:mergeb}
	\end{subfigure}
	\caption{\label{fig:merging}(\subref{fig:mergea}) An example of two colonies merging from experimental images. The two colonies, shown in blue and orange are beginning to merge at 5 days after seeding. The scale bar represents 100$\,\mu$m.  (\subref{fig:mergeb}) Diagram illustrating initially seeded cells and the colonies at time $\tau$, the first time at which the two growing colonies touch each other from a simulation of the cell seeding model. The orange cells are classed as a pair and grow accordingly faster. From \cite{meclonality}.}
\end{figure}

Our multi-population model, Equation~(\ref{eq:model2}), can be implemented to explore the impact of colony growth on clonality \cite{meclonality}. Generating homogeneous populations of clonal cells is of great importance \cite{clonality1,clonality2} as clonally derived stem cell lines maintain pluripotency and proliferative potential for prolonged periods \cite{AMIT2000271}. To achieve this, cross-contamination and merger of colonies (illustrated in Figure~\ref{fig:merging}(a)) should be avoided.

Assuming that, intially, the cells are randomly scattered in a growth area with a particular seeding density (the average number of cells per unit area), each cell (or group of cells) proliferates according to Equation~(\ref{eq:model2}). Each colony is then approximated by a circle, with a certain position in space (the geometric centre of the founding cells) and a radius based on the population size and an assumed cell are of 250\,$\mu$m$^2$ \cite{siriomorphology}. The time at which a colony begins to merge with its neighbour, $\tau$, is the time at which the perfect clonality is lost as illustrated in Figure~\ref{fig:merging}. The simulation leads us to an equation, consistent with experimental data, from which we can estimate the time taken for the first colony merge to occur 
\begin{equation}
\frac{\tau}{1\,\textrm{h}} \approx 100-\frac{n_0}{140\,\textrm{cm}^{-2}},
\end{equation}
\noindent
where $n_0$ is the initial seeding density of cells before their attachment to the substrate in cells/cm$^2$. These results can be used to achieve the best outcome for homogeneous colony growth \textit{in-vitro} by choosing the optimal cell seeding density.

\begin{figure}[h]
	\centering
	\begin{subfigure}[t]{0.38\columnwidth}
		\caption{}
		\includegraphics[width=\columnwidth]{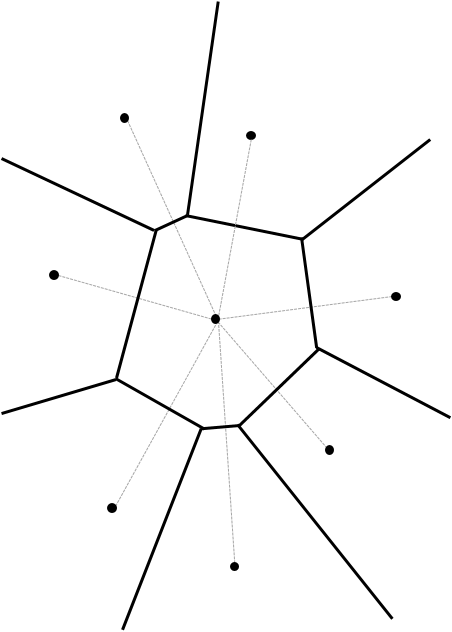}
		\label{fig:pro1}
	\end{subfigure}
	\begin{subfigure}[t]{0.6\columnwidth}
		\caption{}
		\includegraphics[width=\columnwidth]{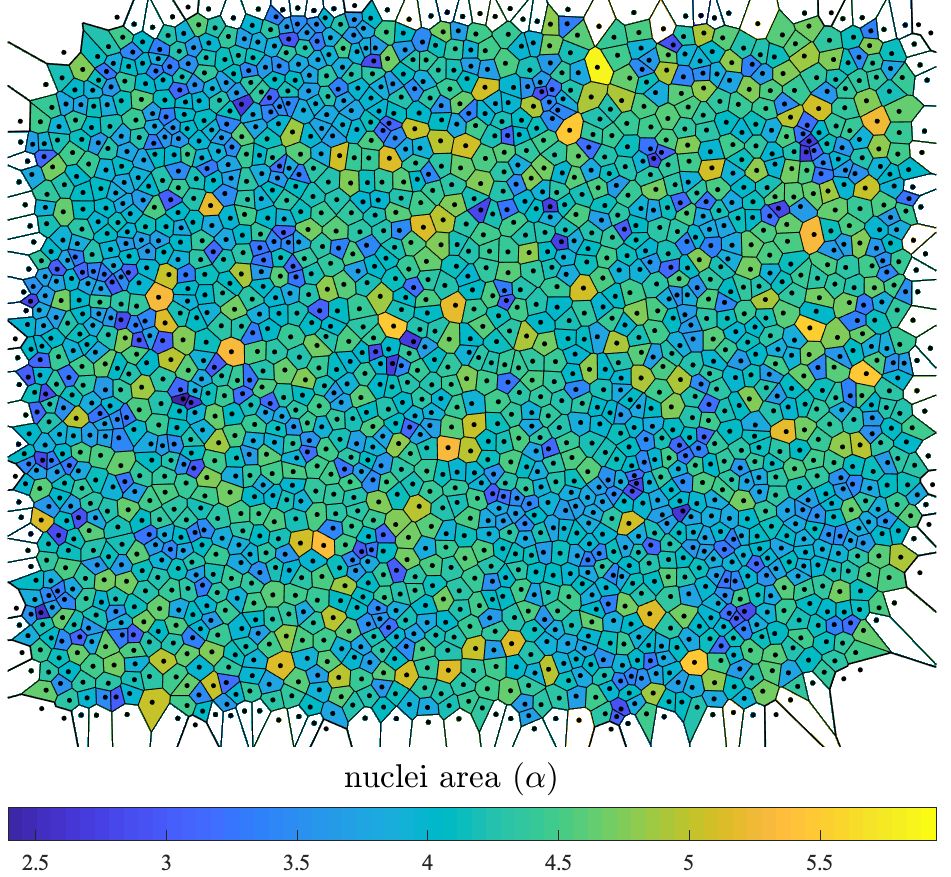}
		\label{fig:pro2}
	\end{subfigure}
\\
	\begin{subfigure}[t]{1\columnwidth}
		\caption{}
		\includegraphics[width=\columnwidth]{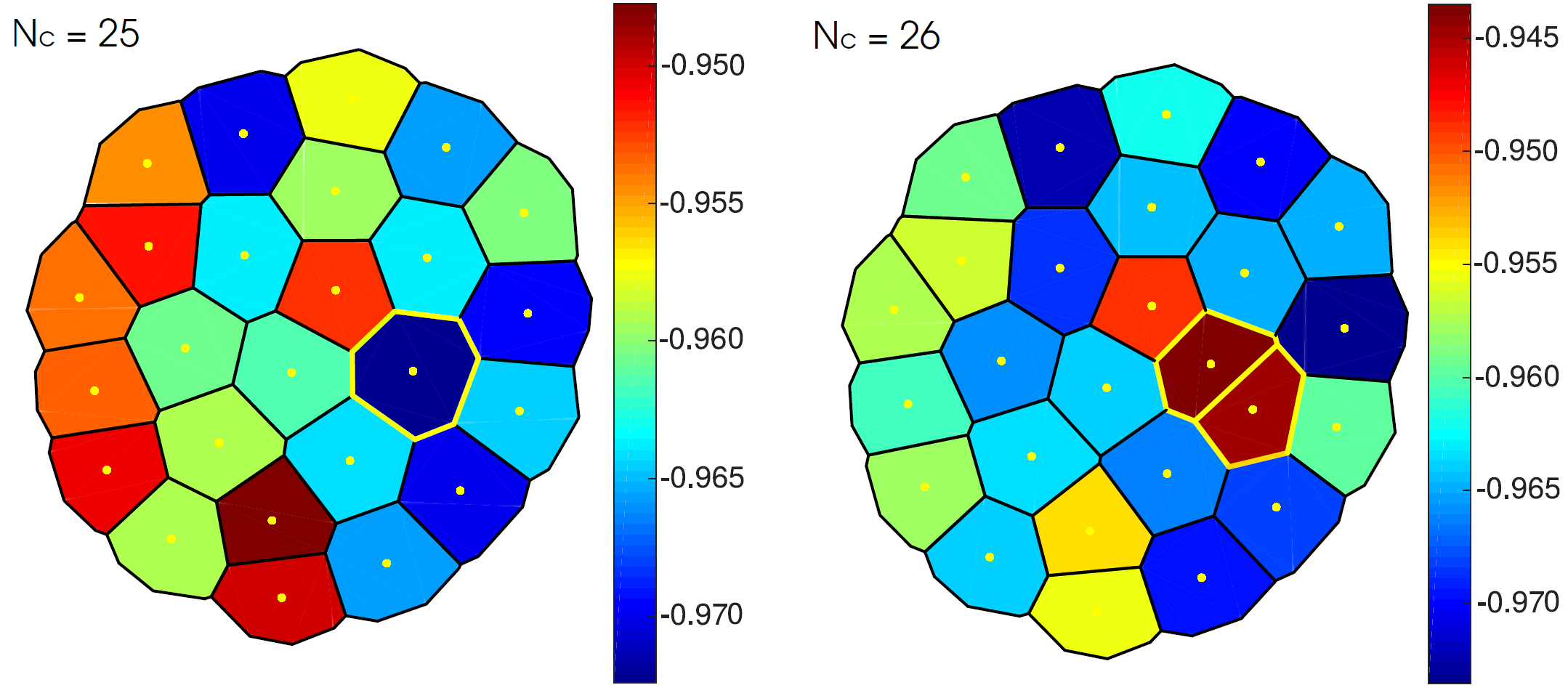}
		\label{fig:pro3}
	\end{subfigure}
	\caption{\label{fig:voronoi}(\subref{fig:pro1}) Voronoi diagram illustrating how colony area is split into tessellated cells. (\subref{fig:pro2}) The Voronoi tessellation obtained from the centroid positions of cells in an experimental microscopic image \cite{siriomorphology}. (\subref{fig:pro3}) Voronoi tessellation to simulate a proliferating hESC colony. The cells divide and give rise to two daughter cells under suitable conditions, see highlighted cells in yellow. Left: the colour bar shows the elastic field in Eq.~\ref{harmonic_potential} with the yellow cell highly stressed due compression from their neighbours. Right: the colour bar shows the same colony with cells coloured according to the stage of their cell cycle, from early in the cell cycle (red) to late (blue).}
\end{figure}

Other spatial models consider each individual cell's position in space. Common vertex based models for adult stem cell proliferation use Voronoi tessellation to describe cell position and areas. The colony area is divided so that the area occupied by a cell is obtained by tracing straight lines between the position of a cell and all its neighbours and drawing a perpendicular line in the middle as shown in Figure~\ref{fig:voronoi}(a). These lines form a convex polyhedron called the Voronoi cell. The Voronoi cells are not uniform in shape and their number of sides varies. The tessellation can be constructed from experimental images using the cell centroid or cell nuclei positions, as shown in Figure~\ref{fig:voronoi}(b) \cite{siriomorphology}. Voronoi tessellation has been used to model adult stem cells in intestinal crypts in 2D \cite{crypts1,Sirio2017} and is now being transferred to hESCs. The model uses an agent-based approximation in which each cell is represented as a Voronoi tessellation of the space \cite{barrio,Sirio2017}. The domain grows according to the pressure flow due to mitotic divisions in the colony. The dynamics between the cells are described by an elastic potential acting on each cell $i$ as
\begin{equation}
V({\bf r}_i, t) = \frac{k_v}{2} \left[ \alpha_i(t) - \overline{\alpha}_0 (t) \right]^2 + \frac{k_c}{2} \left[ {\bf r}_i(t) - {\bf r}_{0i}(t) \right]^2
\label{harmonic_potential}
\end{equation}
\noindent with $k_v$ and $k_c$ elastic constants, $\alpha_i$ the area of each cell, $\overline{\alpha}_0$ the equilibrium area and ${\bf r}_i$ the initial positions of the cells, which do not necessarily correspond to the centroids denoted with ${\bf r}_{0i}$. The first term in the right hand side of Eq.~(\ref{harmonic_potential}) tends to enforce uniform cell size and the second one gives the shape of the cells. Since the forces are conservative, applying the gradient operator to Eq.~(\ref{harmonic_potential}) and adding a drag force, the total force acting on each cell is obtained.

The boundary of the colony is modelled using `ghost cells' whose only function is to bound the domain. Figure~\ref{fig:voronoi}(c) shows a simulated colony undergoing a cell division. Cells in the middle of the colony experience a higher pressure and show mitotic arrest, i.e. they do not divide.

Spatially modelling each individual cell in a colony  in this way raises an important question about the physical process involved in cell division: how does the colony rearrange to make space for new cells? In Voronoi tessellation models \cite{Sirio2017,barrio} the cells re-accommodate themselves according to the potential from the neighbouring cells or the crypt walls. In most square or hexagonal lattice-based models, one daughter cell is placed in the same position as the mother cell while the other is put in a neighbouring position, chosen at random \cite{CAmodelling}, isotropic mitosis. If there is no free position available next to the dividing cell, the neighbouring cells are re-arranged into other available free spaces stochastically until there is a free space next to the dividing cell \cite{NGUYEN2019625} or, if this is impossible, mitosis is suppressed (quiescence) \cite{Khayyeri,PEREZ20072244}. Further experimental time-lapse image data is needed to clarify exactly how the new cells are placed in real colonies.

Proliferation also depends on spatial and environmental factors. There is evidence that high cell density reduces cell proliferation \cite{Wu15}, which has been captured in a model showing preferential cell division at the colony edge \cite{NGUYEN2019625}. Self-organisation of cells has also been observed, where the newly divided (smallest) cells cluster together in patches, separated from larger cells at the final stages of the cell cycle \cite{siriomorphology}. This segregation by cell size allows the interchange of neighbours as the colony grows and could directly influence cell-to-cell interactions and community effects.

Spatial models of hPSCs become increasingly complex with colony size, and it is difficult to successfully incorporate many properties of colony growth along with any collective migratory effects. The question of how colonies re-arrange upon cell divisions requires more experimental investigation to elucidate the best models. The development of these models has already had an impact in understanding the growth of cancer tumours \cite{WANG201570} and wound healing \cite{Tartarini16}. 

\section{Cell pluripotency}
\label{sec:cellpp}

Pluripotency is the defining characteristic of stem cells, often referred to as a cell's `stemness'. It is hPSCs pluripotency that gives them the capability of differentiating into any type of specialised cell in the human body. However, hPSCs can undergo spontaneous differentiation which is undesirable for further experimental applications. Mathematical models of pluripotency are deepening our understanding of how pluripotency is regulated, leading to the optimisation and control of pluripotency in the laboratory.

The decision of a stem cell to remain pluripotent or to differentiate into a particular specialised cell is known as its fate decision. It is not known when a cell makes this decision. Even clonal cells under the same conditions make different fate decisions and it remains unclear how much fate choice is lead by inherited factors versus environmental factors and intracellular signalling. \cite{SYMMONS2016788}. There are several thorough reviews of the computational models of cell fate decisions \cite{cellfate03,cellfate09,cellfate15}. Here we focus on the regulation of pluripotency and spatial patterning within colonies.

Biomedical and clinical applications of hPSC colonies demand tight control of colony pluripotency and homogeneity \cite{Bauwens08}, yet this remains challenging. At a single-cell level, pluripotency is inherently stochastic; indeed, it has been proposed that pluripotency is only defined statistically within a population \cite{MACARTHUR13}. Cells are regulated by their local environment \cite{SHUZUI2019246,Stadhouders19}, notably their beneficial interactions with neighbours \cite{Nemashkalo3042,band}. Colonies exhibit heterogeneous subpopulations of cells with differing levels of PTF expression \cite{ppgenenetwork,Kumar14} suggesting a play-off between disruptive single-cell  and regulatory community effects. Such heterogeneity is undesirable, biasing evolution the trajectories and leading to spatially disordered differentiation \cite{Warmflash14}. Here we will consider  intra-cellular models of pluripotency based on PTFs, and the spatial organisation of pluripotency at the colony level.  

\subsection{Fluctuating PTFs}

The positive-feedback regulation between PTFs (the transciption factors which regulate pluripotency, see Section 2.1) was first described as a first order differential equation model using the Hill equations \cite{Chickarmane06}. However, the parameters of such a model are difficult to estimate accurately \cite{Gutenkunst07}. More recently, PTFs have been modelled through branching processes \cite{Greaves17}. A  thorough review of the models of pluripotency is available \cite{ppmodels}, along with a review of computational modelling of the fate control of mouse embryonic stem cells, with many models transferable to hPSCs \cite{cellfate15}.

\begin{figure}[h]
	\centering
	\begin{subfigure}[t]{.9\columnwidth}
		\caption{}
		\includegraphics[width=\columnwidth]{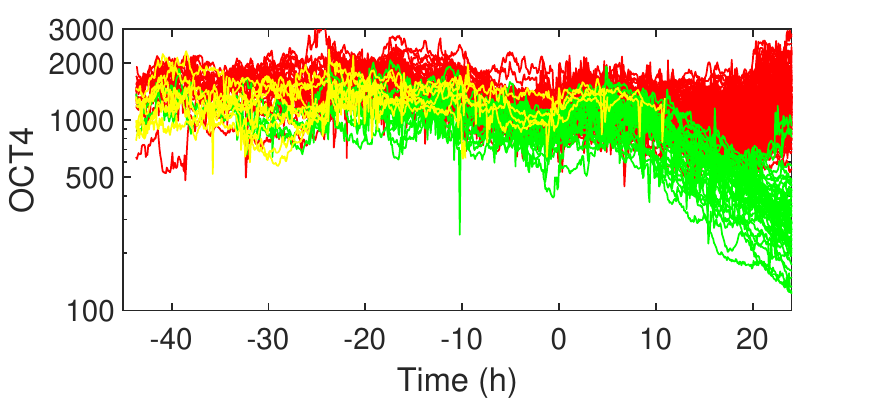}
		\label{fig:wolffa}
	\end{subfigure}
	\\
	\begin{subfigure}[t]{1\columnwidth}
		\caption{}
		\includegraphics[width=\columnwidth]{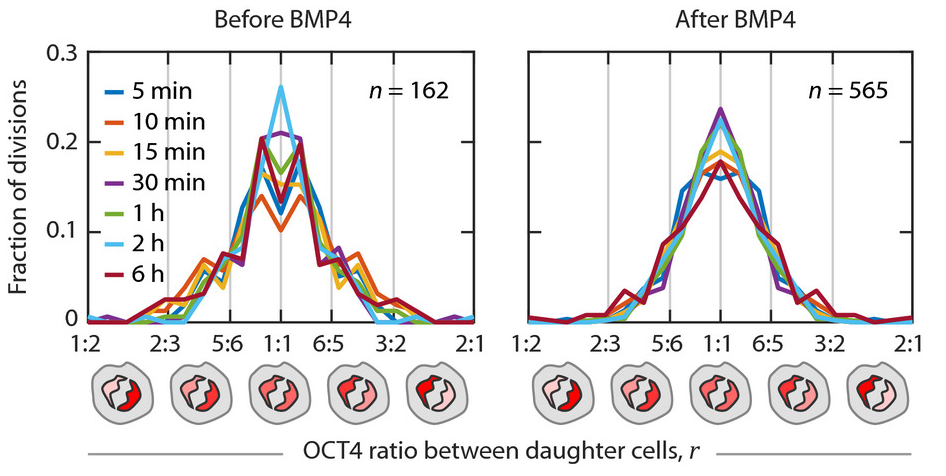}
		\label{fig:wolffb}
	\end{subfigure}
	\caption{\label{fig:Wolff}(\subref{fig:wolffa}) OCT4 values over time, coloured by cell fate - pluripotent cells (red), unknown (yellow) and differentiated (green). Time zero is the time the BMP4 is added to the cells. Figure reproduced from \cite{Wolff}. (\subref{fig:wolffb}) The OCT4 splitting ratio between daughter cells before and after BMP4 addition. Figure from \cite{Wolff}.}
\end{figure}

Recent experimental work has investigated how the PTFs vary over time, and how maternal PTFs are transmitted and distributed between the daughter cells \cite{Wolff}. The OCT4 abundance in the cells was tracked over time before and after the addition of an agent which induces differentiation (BMP4). The cell fates were also recorded. The OCT4 values over time for all cells, organised by cell fate (pluripotent, unknown or differentiated), are shown in Figure~\ref{fig:Wolff}(a).

We are currently working on modelling the trends and fluctuations in pluripotency over time based on the experimental OCT4 data in \cite{Wolff}. First we quantified the nature of the persistence of the OCT4 time series. The Hurst exponent, $H$ is a measure of the the long-term memory of a time series, with $H=0.5$ corresponding to Brownian motion, $0<H<0.5$ anti-persistence (a preference to change the direction of the last step) and $0.5<H<1$ persistence (a preference to continue the trend of the last step). The mean Hurst exponent for the OCT4 data is 0.36, signifying anti-persistence and importantly suggesting self-regulation of pluripotency. We are exploring stochastic modelling techniques, particularly fractional Brownian motion to capture the anti-persistence and the stochastic logistic equation to model the evolutions of the cells. Both of these models are well established, however their application to modelling pluripotency is novel.

As the general OCT4 levels is inherited after cell division, pluripotency levels are most similar among closely related cells even when a reasonable level of randomness is allowed for \cite{Wolff}. The analysis in \cite{Wolff} also shows that OCT4 is not always equally allocated between daughter cells upon cell division with the split being sometimes asymmetric, as shown in Figure~\ref{fig:Wolff}(b). Models of pluripotency inheritance should take into account this variation in the splitting ratio upon cell division. This study also suggests that a cell's decision to differentiate is largely determined before the differentiation stimulus is added and can be predicted by a cell's pre-existing OCT4 signalling patterns. These results imply that the choice between developmental cell fates can be largely predetermined at the time of cell birth through inheritance of a pluripotency factor \cite{Wolff}.

These results highlight the important properties for models of hPSC pluripotency to capture at the individual cell level: the stochastic inheritance of PTFs, the anti-persistence or self-regulation of pluripotency and the pre-determined cell fate decision. Suitable models can then be developed to not only represent the behaviour on a individual cell scale, but also the colony scale.

\subsection{Spatial organisation}

\begin{figure}
	\includegraphics[width=1\columnwidth]{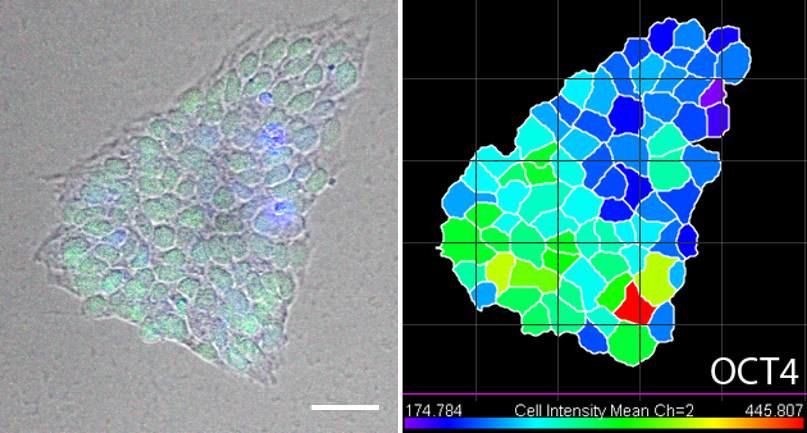}
	\caption{A microscopy image of a hESC colony at 72\,h after seeding, alongside a colour-coded version of the same colony quantifying the level of expression of OCT4. Red represents the highest pluripotency with blue representing the lowest. Scale bar represents $50\,\mu$m.}
	\label{fig:IMARIS}       
\end{figure}

Pluripotency also shows spatial variation on the colony scale. Preliminary experiments monitoring the OCT4 levels in colonies grown from single cells at 72\,h post seeding show that pluripotency is clustered, with highly pluripotent cells grouped together, as shown in Figure~\ref{fig:IMARIS}.

\begin{figure}
	\centering
\begin{subfigure}[t]{.49\columnwidth}
	\caption{}
	\includegraphics[width=\columnwidth]{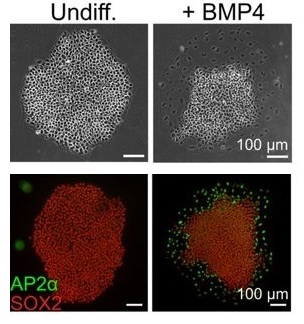}
	\label{fig:banda}
\end{subfigure}
\begin{subfigure}[t]{.49\columnwidth}
	\caption{}
	\includegraphics[width=\columnwidth]{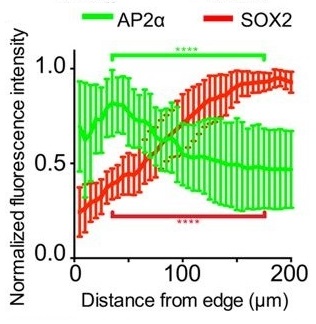}
	\label{fig:bandb}
\end{subfigure}
	\caption{(\subref{fig:banda}) Phase (top) and immunostaining images (bottom) of hESC colonies before and after BMP4 addition. (\subref{fig:bandb}) Analysis of expression of a pluripotency marker SOX2 and differentiation marker (AP2$\alpha$) 3 days after BMP4 treatment. Fluorescent intensity is plotted as a function of distance from the colony edge and normalized to the maximum intensity of each colony [n= 20 colonies, $p< 0.0001$ and represents statistics for AP2$\alpha$ (green) and SOX2 (red) levels between distance 35\,$\mu$m and 175\,$\mu$m from the edge using a two-tailed paired t-test]. Error bars represent standard deviations from the mean. Adapted from \cite{band}.}
	\label{fig:band}       
\end{figure}

The differentiation of hPSCs also shows distinctive spatial patterning \cite{band,Warmflash14}. Experiments monitoring the pluripotency marker SOX2 and the differentiation marker AP2$\alpha$ have shown that differentiation occurs preferentially at the colony periphery in a band of constant width, independent of colony size, as illustrated schematically in Figure~\ref{fig:cellscales}(c) and shown in Figure~\ref{fig:band} \cite{band}. These differentiated cells originate from the outer third of the colony, and remain at the edge. This provides important information for modelling the spatial patterning of the pluripotent state.

This within-colony spatial patterning behaviour of the differentiation has been captured by a mechanical bidomain model \cite{Auddya_2017}, a continuum model first developed to describe the elastic behaviour of the cardiac tissue \cite{Bidomainheart}. The model predicts that differentiation and traction forces occur within a few length constants of the colonies edge, consistent with the experimental results for differentiation in hPSCs \cite{band,Warmflash14}. The model assumes that differences in displacement are responsible for any mechanotransduction (chemical processes through which cells sense and respond to mechanical stimuli) and describes both the intra and extra-cellular spaces in colonies with relationships between stress, strain and pressure forces. The basic equation for the difference between the intra and extra-cellular displacements for changing distance from the colony centre $r$, $u_r$ and $w_r$ respectively as
\begin{equation}
u_r-w_r=-\frac{T\sigma}{4\nu} \rm{exp}\left\{\frac{r-R}{\sigma}\right\},
\end{equation}
where $T$ is a uniform stress caused by the growth and crowding of cells, $\nu$ is the shear modulus, $\sigma$ is a length constant and $R$ is the colony radius. This model shows that if the difference between the intra-cellular and extra-cellular displacements drives the differentiation, then differentiation is confined to the edge of the colony. This model could be further developed to include more complicated geometries as currently the colony is assumed to be circular to allow analytical solutions to the model equations. Furthermore, it is worth investigating whether the cell growth represented by the tension $T$ is a function of $u_r-w_r$ alone, as observations for hESCs suggest distinct actin organization and greater myosin activity near the colony edge, implying that $T$ could be non-uniform \cite{band}.

Further experiments are needed to collect data on the pluripotency of cells across colonies. Analysis of the data using techniques common in spatial statistics will allow the continued development of pluripotency models on the colony scale.

\section{Discussion}
\label{sec:disc}

Mathematical and computational models of hPSC growth are essential in formulating non-invasive predictive tools. Although we have focussed on hPSCs here, it is worth noting that similar models are used to describe the reprogramming of somatic cells into iPSCs, which is still a low-yield process with the underlying processes of cell fate decision uncharacterised \cite{TAKAHASHI2007861}. As the reprogramming is a stochastic process, most mathematical models in this area probabilistic \cite{Hanna09}. A model describing cell types as a set of hierarchically related dynamical attractors representing cell cycles has lead to the identifications of two mechanisms for reprogramming in a two-level hierarchy: cycle-specific perturbations and a noise-induced switching \cite{Hannam_2017}. These reprogramming protocols make specific predictions concerning reprogramming dynamics which are broadly in line with experimental findings. Another reprogramming model using a two-type continuous-time Markov process with a constant reprogramming rate has revealed two different modes of cellular reprogramming dynamics: TF expression alone leads to heterogeneous reprogramming while TFs plus certain other factors homogenise the dynamics \cite{reprogrammingmodel}.

Here we have discussed some key properties of hPSCs: cell kinematics, cell proliferation and cell pluripotency. However, there are other important factors which could be included in modelling, e.g., environmental factors, cell-cell signalling, intra-cellular properties and collective migration. Models isolating a few of these key properties have often captured experimental results well. For example, focussed migration models have lead to a greater understanding of the behaviour of isolated cells \cite{Li,me2,me} and the movement of cells within colonies \cite{NGUYEN2019625,Hoffman}. There are many population models for colony proliferation, taking into account cell divisions and deaths, providing a distinct computational advantage over more complex spatio-temporal models. Models of colony growth have been used to investigate the impact of colony expansion on clonality \cite{meclonality}, cell regeneration within intestinal crypts \cite{crypts1,Sirio2017} and tumour growth \cite{WANG201570}.

Many current efforts focus on modelling cell pluripotency and cell fate, as applications of hPSCs require greater control over pluripotency and differentiation trajectories. The stochastic nature of pluripotency at the single cell level \cite{MACARTHUR13}, along with regulatory community effects leads to heterogeneous sub-populations across colonies \cite{ppgenenetwork,Kumar14}. Recent studies of the fluctuations of PTFs throughout colonies \cite{Wolff} and spatial patterning of differentiation \cite{band,Warmflash14} are being used to inform the development of models of pluripotency and cell fate.

Developing comprehensive models of hPSCs remains challenging, due to their many complex properties across multiple scales, and not yet characterised collective behaviour effects. It is also difficult to match parameters with experimental observations. Model refinement should be based on a two-way interaction with experiments; model parameters should be informed by experimental results, and models should influence experimental design. Such models have already helped provide an insight into tissue formation, wound healing, tumour growth and the reprogramming of iPSCs and will no doubt continue to do so as these models progress. 

%
%

\begin{acknowledgements}
We acknowledge financial support from Newcastle University, and European Community 
(IMI-STEMBANCC, IMI-EBISC, ERC \#614620), NC3R NC/CO16206/1) and BBSRC UK (BB/I020209/1) for providing financial support for this work. IN acknowledges the grant from the Russian Government Program for the recruitment of the leading scientists into Russian Institution of Higher Education 14.w03.31.0029. SOF thanks the National Council for Science and Technology (CONACYT), Mexico, for the scholarship CVU-174695. AS acknowledges partial financial support of the Leverhulme Trust (Grant RPG-2014-427).
\end{acknowledgements}

\section*{Conflict of interest}
The authors declare that they have no conflict of interest.

\bibliographystyle{unsrt}      
\bibliography{mybib}   

\end{document}